\documentclass[prl,twocolumn,showpacs]{revtex4}

\usepackage{graphicx}% Include figure files

\begin{document}
\title{Transient structures in a granular gas}
%Clustering and vortex formation as a transient process in Granular Gas}
% of viscoelastic particles}

%\title{Do clusters exist in realistic Granular Gases? \\
%       Do clusters exist in Granular Gases of viscoelastic particles? \\
%       Clustering and vortex formation as a transient process in realistic Granular Gases}
\author{Nikolai Brilliantov$^{1,2}$, Clara Salue\~na$^{3}$, Thomas Schwager$^{1}$ and Thorsten P\"{o}schel$^1$}
\affiliation{$^1$Institut f\"ur Biochemie, Charit\'e, Monbijoustr. 2, 10117
Berlin, Germany}
\affiliation{$^2$Moscow State University, 119899 Moscow, Russia}
\affiliation{$^3$Universitat Rovira i Virgili, 43007 Tarragona, Spain}

\date{\today}

\begin{abstract}
A force-free granular gas is considered with impact-velocity
dependent coefficient of restitution as it follows from the model of
viscoelastic particles. We analyze structure formation in this
system by means of three independent methods: Molecular Dynamics,
numerical solution of the hydrodynamic equations and linear
stability analysis of these equations. All these approaches
indicate that structure formation occurs in force-free granular
gases only as a transient process.
\end{abstract}

%45.70.-n             Granular systems
%45.70.Qj             Pattern formation (granular systems)
%47.20.-k             Hydrodynamic stability

\maketitle Cluster and vortex formation in force-free granular
gases is a most striking phenomenon which makes these systems so
distinct from gases of {\em elastic} particles, like molecular
gases. First detected and explained by Goldhirsch and Zanetti
\cite{GoldhirschZanetti:1993} and McNamara \cite{McNamara:1993},
clustering and later vortex formation \cite{BritoErnst:1998} have
been intensively studied (e.g.
\cite{BritoErnst:1998,BreyRuizMonteroCubero:1999,NoijeErnst:2000}).
Clustering has been also reported in recent simulations of the
hydrodynamic equations \cite{HillMazenko:2002}. In all these
studies the simplifying assumption of a constant coefficient of
restitution $\varepsilon$ has been used.

It is known, however, that the coefficient of restitution is not a
material constant, but a function of the impact velocity, which
reads for the most  simple model of  viscoelastic spheres
\cite{Ramirez:1999}:
\begin{equation}
 \label{epsilon} \varepsilon (g)=1-\gamma \left| g \right|^{1/5} + (3/5) \gamma^2 \left|g \right|^{2/5} \mp \cdots
\end{equation}
Here $g$ is the normal component of the impact velocity, and
$\gamma$ is a known function of the particles' material properties
\cite{Ramirez:1999}. The typical value of $\varepsilon$,
corresponding  to the thermal velocity $v_T=\sqrt{2T(t)/m}$
($T(t)$ is the temperature of the gas and  $m=1$ is the particle mass)
behaves as $\varepsilon_{T} \sim 1- \gamma T^{1/10}$, i.e., it
tends to the elastic limit, $\varepsilon  \to 1 $,  as the gas
cools down, $T \to 0$.

%%%%%%%%%%%%%%%%%%%%%%%%%%%%%%%%%%%%%%%%%%  ^^^^^^^^^^ %%%%%%%%%%%%%%%%%%%%%
Since a force-free gas of elastic particles tends to be
homogeneous, one can naively assume that a granular gas of
viscoelastic particles tends to be finally uniform as well.
However, this is not necessarily true: The collisions become
perfectly elastic only in the limit $g=0$ when all particles are
at rest. If the gas cools down too fast the residual structures
may get frozen and persist due to a lack of kinetic energy. We
illustrate this for a more general model   of $\varepsilon(g)$,
with $(1-\varepsilon_{T}^2) \sim T^{\beta}$. The cooling rate of
such gas is estimated as $\dot {T} \sim - n
(1-\varepsilon_{T}^2)T^{3/2} \sim - n T^{3/2+\beta}$
\cite{GoldhirschZanetti:1993,BrilliantovPoeschel:2000visc}, where
$n$ is the gas number density. The gas density decreases with time
due to cluster growth. If we assume  $n \sim t^{-\alpha }$
($0<\alpha <1$) \cite{remarkalpha}, we obtain the estimate for the
gas temperature, $T\sim t^{-z}$, $z=2(1-\alpha)/(1+2\beta)$,  and
for the gas pressure, $P=nT\sim t^{-\alpha-z}$. If $T_{\rm cl}$
and $n_{\rm cl}$ are respectively the temperature and number
density inside the clusters, similar estimates yield: $\dot
{T}_{\rm cl}  \sim - n_{\rm cl} T_{\rm cl}^{3/2+\beta}$. Assuming
that $n_{\rm cl}$ keeps approximately constant, we obtain $T_{\rm
cl} \sim t^{-2/(1+2 \beta)}$ and estimate the pressure in
clusters, $P_{\rm cl} \sim f(n_{\rm cl})T_{\rm cl} \sim t^{-2/(1+2
\beta)}$, where the factor $f$ does not depend on $T$. The
necessary condition for clusters to dissolve reads $P_{\rm cl} > P
$ for $t \to \infty$, i.e., $\alpha+z
> 2/(1+2 \beta)$. With $z $ given above this is equivalent to
the condition $\beta > 1/2$, that is, only gases with a pronounced
dependence of $\varepsilon_{T}$ on $T$ evolve to a uniform final
state. For gases  of viscoelastic particles $\beta =1/10 <1/2$,
hence it is not clear whether structures arise only temporarily in
these systems, or appear and get frozen.

%%%%%%%%%%%%%%%%%%%%%%%%%%%%%%%%%%%% vvvvvvvvvvvvvv %%%%%%%%%%%%%%%%%%%%%%%%%%%%%%%%%%%%%

To address this problem we study the evolution of a gas of
viscoelastic particles by means of three independent methods: event-driven
Molecular Dynamics (MD), numerical solution of the hydrodynamic equations and
linear stability analysis of the HD equations.

\begin{figure}[t!]
  \centerline{\includegraphics[width=8.7cm]{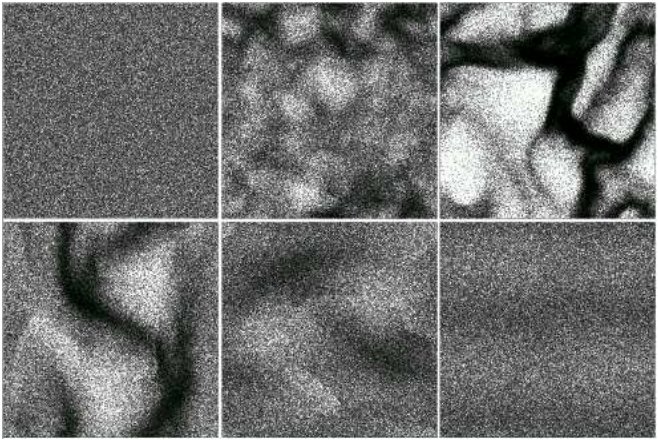}}
%  \centerline{
%    \includegraphics[width=2.8cm]{snap0.eps}\hspace{-0.05cm}
%    \includegraphics[width=2.8cm]{snap2.eps}\hspace{-0.05cm}
%    \includegraphics[width=2.8cm]{snap8.eps}
%    }
%  \centerline{
%    \includegraphics[width=2.8cm]{snap75.eps}\hspace{-0.05cm}
%    \includegraphics[width=2.8cm]{snap200.eps}\hspace{-0.05cm}
%    \includegraphics[width=2.8cm]{snap700.eps}
%    }
%\centerline{\includegraphics[width=8.5cm]{allsnapsvis.ps}}
  \caption{MD simulation of $N=10^5$ viscoelastic particles. Clusters appear as transient structures. The snapshots where taken after 0, 200, 800, 7,500, 20,000 and 70,000 collisions per particle.}
  \label{MDclusters}
\end{figure}

\begin{figure}[t!]
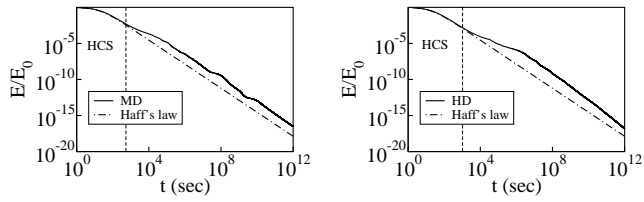

% \centerline{\includegraphics[width=8.5cm]{}}
\includegraphics[width=4cm]{energyMDneu.eps}\hspace{0.3cm}
\includegraphics[width=4cm]{meant_contribL2.eps}
\caption{ The energy decay as obtained by MD (left) and by the
numerical solution of   the HD equations (right). The dash-dotted
lines show the modified Haff law, Eq. (\ref{T_HCS}). The end of
the HCS is indicated by the vertical dashed lines.}
  \label{MDHYDHaff}
\end{figure}

The MD simulations with periodic boundary conditions were
performed for a 2d gas of $N=10^5$ particles of diameter
$\sigma=1$, with the coefficient of restitution according to Eq. (\ref{epsilon}).
The gas has the packing fraction $\phi=n \pi \sigma^2/4=0.2$,
initial temperature $T_0=1$ and the dissipative coefficient
$\gamma =0.0577$ which corresponds to the initial coefficient of restitution
$\varepsilon_{T} \sim 1- \gamma \simeq 0.94$.
Starting the simulations with homogeneous distribution of
particles (homogeneous cooling state, HCS), clusters appear and
grow until they reach the {\em system size}, then the clusters
dissolve (Fig. \ref{MDclusters}).
Figure  \ref{MDHYDHaff} shows the evolution of the average particle energy, $E(t)=(1/2N) \sum_i m v_i^2 $, which evolves in the HCS according to the modified  Haff law
  \cite{BrilliantovPoeschel:2000visc},
\begin{equation}
\label{T_HCS} E(t)/E_0 = T_{\rm h}(t)/T_0 =\left( 1 + t/\tau_0 \right)^{-5/3} \, ,
\end{equation}
where
\begin{equation}
\label{eq:deftau0} \tau_0^{-1}=(24/5)q_0 \tau_c^{-1}(0)  \delta \, ,~~ \delta=(\gamma/C_1)(T_0/m)^{1/10} \,.
\end{equation}
Here $\tau_{c}^{-1}(t)=2n \sigma g_2(\sigma) \sqrt{\pi T(t)/m}$ is
the mean collision time, $g_2(\sigma)=(1-7\phi/16)/(1-\phi)^2$ is
the contact value of the  pair correlation function. The constants
$q_0 \approx 0.173$ and $C_1 \approx 1.1534$ are also known
analytically  \cite{BrilliantovPoeschel:2000visc}.  At later times
the deviation from the Haff law becomes pronounced, however, as the
gas further evolves the clusters and vortices dissolve, and the
system approaches the regime of the homogeneous cooling, see
Fig. \ref{MDclusters}.

\vspace{0.1cm}

%%%%%%%%%%%%%%%%%%%%%%%%%%%%%%%%%
To explain the observed effects we consider the hydrodynamic
equations
\cite{BrilliantovPoeschel:2001roy,BrilliantovPoeschel:2003,HillMazenko:2002}
for the number density $n(\vec{r},t)$, flow velocity
$\vec{u}(\vec{r},t)$ and temperature field $T(\vec{r},t)$
\begin{eqnarray}
\label{hydron}
&& \frac{\partial n}{\partial t}    = - \nabla_i \left( n u_i \right)  \\
\label{hydrou}
&& \frac{\partial u_i}{\partial t} = - \left( u_j\nabla_j \right)u_i +\frac{1}{nm}\nabla_j
   \left( \eta_{ijkl} \nabla_k u_l -P \delta_{ij} \right) \\
&& \label{hydroT}
   \frac{\partial T}{\partial t}   = - \left( u_j\nabla_j \right) T -\frac{P}{n}  \left( \nabla_i u_i \right)
   + \frac{1}{n} \eta_{ijkl}\left( \nabla_k u_l \right)\left( \nabla_j u_i \right) \nonumber \\
&& \, \, \, \, \, \, \, \, \, \, \, \,
   +\frac{1}{n} \nabla_i \left( \kappa \nabla_i T \right) + \frac{1}{n} \nabla_i \left( \mu  \nabla_i n  \right)
   - \zeta T \, .
\end{eqnarray}
Here $P=nT[1+(1+\varepsilon)\phi g_2(\sigma)]$ and
$\eta_{ijkl}=\eta ( \delta_{ik}\delta_{jl} +
\delta_{il}\delta_{jk} - \delta_{ij} \delta_{kl})$ is the
viscosity tensor.  $\eta$ and $\kappa$ respectively are the
coefficients of viscosity and thermal conductivity; $\zeta$ is the
cooling coefficient and the coefficient $\mu$ is specific for
granular gases \cite{SelaGoldhirsch:1998,BreyDuftyKimSantos:1998}.
The coefficients $\eta$, $\kappa$, $\mu$ and $\zeta$ have been
recently derived for a  gas of viscoelastic particles
\cite{BrilliantovPoeschel:2003}. They may be written as an
expansion ($b=\eta, \, \kappa, \, \mu, \, \zeta$),

\begin{equation}
\label{etakapmu}
b=b_0+b_1 \delta^{\prime} + b_2 \delta^{\prime \, 2} \,,
\end{equation}
where $\delta^{\, \prime} (t) = \delta \left[2T(t)/T_0
\right]^{1/10}$ is the time-dependent dissipation parameter and
the microscopic expressions for $b_{0}$, $b_{1}$, $b_{2}$ are
given in \cite{BrilliantovPoeschel:2003}. We wish to stress that
the temperature dependence of these coefficients differs
drastically from the case $\varepsilon = {\rm const}$, see e.g.
\cite{LunSavageJeffreyChepurniy:1984,SelaGoldhirsch:1998,BreyDuftyKimSantos:1998}.

%%%%%%%%%%%%%%%%%% Hydrodinamics simulations %%%%%%%%%%%%%%%%%%

We numerically solve the hydrodynamic equations
(\ref{hydron},\ref{hydrou},\ref{hydroT}) with the coefficients Eq.
(\ref{etakapmu}) calculated for the same microscopic parameters
$\gamma$, $\sigma$, $m$ as used for the MD simulations. We use a
finite volume discretization scheme of global second order on a
staggered grid. The integration in time is done through a TVD
multi-step scheme of fourth order \cite{ShuTVD}. We use a
50$\times$50 computational domain with periodic boundary
conditions. Special care has been taken treating the  advection
terms in (\ref{hydron},\ref{hydrou},\ref{hydroT}) \cite{limiter}.
%. Due to the
%final dissolution of the clusters, a simple QUICK scheme is
%capable to sustain the computation without ever reaching the
%problematic limit $n \to 0$ in the case of small systems, but for
%larger systems that becomes a critical issue. To improve the
%robustness of the algorithm, a limiter has been suplemented to the
%base third order upwinding scheme (QUICKEST), which preserves
%monotonicity and minimizes oscillations at the discontinuities
%\cite{limiter}.

\begin{figure}[t!]
  \centerline{\includegraphics[width=8.7cm]{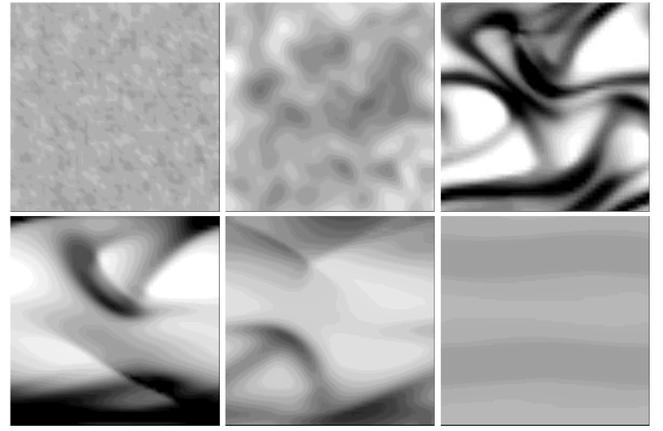}}
%  \centerline{
%    \includegraphics[width=2.8cm]{NEW/tau5.eps}\hspace{-0.08cm}
%    \includegraphics[width=2.8cm]{NEW/tau200.eps}\hspace{-0.08cm}
%    \includegraphics[width=2.8cm]{NEW/tau1250.eps}
%    }

%  \vspace{-0.01cm}
%  \centerline{
%    \includegraphics[width=2.8cm]{NEW/tau7500.eps}\hspace{-0.08cm}
%    \includegraphics[width=2.8cm]{NEW/tau35000.eps}\hspace{-0.08cm}
%    \includegraphics[width=2.8cm]{NEW/tau60000.eps}
%    }
  \caption{ The gas density obtained from the  numerical solution of the HD equations (\ref{hydron},\ref{hydrou},\ref{hydroT}) with periodic boundary and random initial conditions. The parameters $\gamma, \phi, T_0$ are the same as in the MD simulations. Density inhomogeneities (clusters) appear but eventually dissolve. The snapshots were taken after
%400,  600, 1,000, 1,200, 7,500 and 45,000
0, 200, 800, 7,500, 20,000 and 70,000
 collisions per particle, in correspondece to the snapshots in Fig. \ref{MDclusters}. The number of collisions was computed using the average temperature.}
  \label{HYDclusters}
\end{figure}

We start the numerical integration with random initial conditions
for the density and flow-velocity field (thermal fluctuations) and
confirm the transient character of the pattern formation, Fig.
\ref{HYDclusters}.  The cooling curve for $E(t)$ \cite{energy}
demonstrates qualitatively the same behavior as observed in the MD
simulations, Fig. \ref{MDHYDHaff} (right).

\begin{figure}[t!]
% \centerline{\includegraphics[width=8.5cm]{}}
\includegraphics[width=8cm,clip]{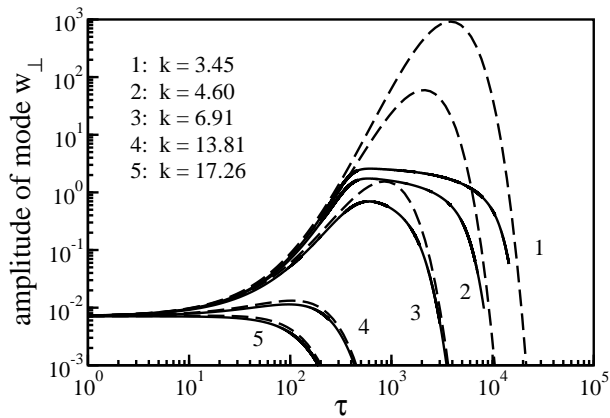}
  \caption{Evolution of the reduced amplitude of the shear mode, $ w_{k, \, \perp}(\tau) \equiv u_{k \, y}(\tau)/v_T(\tau)$.
 Initially only a single shear mode with $k=2 \pi /L$ and
  with an amplitude of the order of the thermal fluctuations is excited.
  Full lines -- numerical solution of the hydrodynamic equations, dashed
  lines -- results of the linear analysis, Eq. (\ref{eq:linstabperp}).
  The parameters are the same as in  Fig. \ref{HYDclusters}.
  Dimensionless time and length are used (see text). Note that the
  different $k$ correspond to a different system size.}
  \label{MDHYDTrans}
\end{figure}

\begin{figure}[t!]
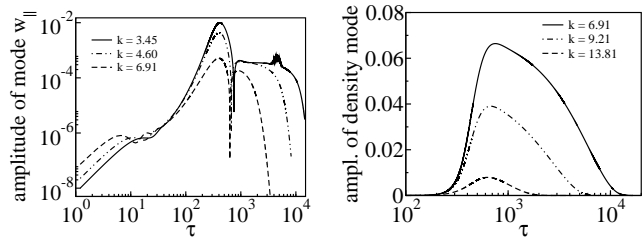

% \centerline{\includegraphics[width=8.5cm]{}}
\includegraphics[width=4cm]{vxfft.eps}\hspace{0.3cm}
\includegraphics[width=4cm]{rhofft.eps}
 \caption{Evolution of the reduced amplitude of the longitudinal mode, $ w_{k, \, ||}(\tau) \equiv u_{k \, x}(\tau)/v_T(\tau)$
 (left) and of the density mode $\rho_k(\tau)$ (right). Parameters, units and initial conditions are the same as in Fig. \ref{MDHYDTrans}.}
  \label{HYDparden}
\end{figure}

To study the mechanism of pattern evolution more directly we
considered initial conditions with a superimposed sinusoidal mode
and observed similar transient structures. For reasons explained
below, we have chosen the transverse velocity (shear) mode. It has
the components, $u_y(\vec{r},0)=u_{k \, y}(0) \sin (k\,x)$,
$u_x(\vec{r},0)=0$, $T(\vec{r},0)=T_0$, $n(\vec{r},0)=n_h$, where
$k=(2\pi/L)l$, ($l=1,2\dots$) is the wave-number, $L$ is the
system size and $u_{k \, y}(0)$ is the initial amplitude of the
mode. Solving the hydrodynamic equations numerically we analyze
the evolution  of the shear mode, Fig. \ref{MDHYDTrans},
the longitudinal mode and  the density mode, Fig. \ref{HYDparden}.

%\vspace{0.2cm}

As it follows from Fig. \ref{MDHYDTrans}, the shear mode (in
reduced units) is unstable for wave-vectors smaller than a
threshold value.  This instability leads to formation of vortices
\cite{BritoErnst:1998,BreyRuizMonteroCubero:1999}. The growing
shear mode initiates the growth of the mode of longitudinal
velocity which, in its turn, causes a  growth of the density mode
corresponding to clustering, see Fig. \ref{HYDparden}. This
indicates that clustering is  driven mainly by a nonlinear
mechanism
\cite{GoldhirschZanetti:1993,BreyRuizMonteroCubero:1999}: As
explained below, a single {\em transverse}  mode, cannot excite by
a linear mechanism, neither a longitudinal mode, nor a density
mode. Hence only nonlinear coupling between the shear and
longitudinal modes may cause the observed excitation of the
longitudinal  mode \cite{secpeak}, which further initiates the
growth of the density mode, i.e. clustering. Similar mechanisms
govern the initial stage of pattern formation in a gas with
$\varepsilon = {\rm const}$
\cite{GoldhirschZanetti:1993,BreyRuizMonteroCubero:1999}. Contrary
to the case $\varepsilon = {\rm const}$, in a force-free gas of
viscoelastic particles all modes eventually decay, and only
transient structures appear.

To obtain quantitative estimates we perform a stability analysis of the HD equations (\ref{hydron},\ref{hydrou},\ref{hydroT}) with respect to the HCS at density $n_{\rm h}$, and temperature $T_{\rm h}(t)$. We assume that the deviations $ \delta T(\vec{r}, t) = T(\vec{r}, t) - T_{\rm h}(t)$ and  $\delta n(\vec{r}, t) = n(\vec{r}, t) - n_{\rm h}$ are small,
\begin{equation}
\theta \equiv \frac{\delta T }{T_{\rm h}} \ll 1 \, , \qquad \rho \equiv  \frac{\delta n }{n_{\rm h}} \ll 1 \, ,
\qquad | \vec{w} | \equiv \frac{ |\vec{u}| }{v_T} \ll 1
\end{equation}
and linearize these  equations. Below we will use the dimensionless time $\tau$, measured in units of  $\tau_c(t)/2$, and  dimensionless  length $\vec{\hat{r}}$, measured in units of $l_0/2$ ($l_0^{-1}= \sqrt{2 \pi} n \sigma g_2(\sigma)$ is the mean free path) and write the linearized equations for the Fourier transforms of the fields $\theta(\vec{r}, t)$, $\rho(\vec{r})$ and
$\vec{w}(\vec{r}, t)$:
\begin{eqnarray}
\label{Fmodes1} &&\! \! \frac{\partial \vec{w}_{\vec{k}\perp}}{\partial\tau} =\left[\beta \,  \delta^{\prime}(
\tau ) - k^2 \right]
\vec{w}_{\vec{k}\perp} \\
\label{Fmodes2}
&&\! \!\frac{\partial \rho_{\vec{k}}}{\partial \tau} = - i k  w_{\vec{k}
||} \\
\label{Fmodes3} &&\! \!\frac{\partial \theta_{\vec{k}}}{\partial\tau} = -\left[4 \tilde{\mu}_1 k^2 + 2\beta \,
\right]
\delta^{\prime}( \tau ) \rho_{\vec{k}} \nonumber \\
&&\! \!  \, \, \, \, \, \, \, \, \, \, \, \, \, \, \, \, \, \, \, - \left[4 k^2 + (6\beta/5)\,  \delta^{\prime}(
\tau )\right] \theta_{\vec{k}}
- i k  w_{\vec{k} ||} \\
\label{Fmodes4} &&\! \! \frac{\partial w_{\vec{k} ||}}{\partial\tau} = \! -\frac12  ik \rho_{\vec{k}} -\frac12  ik
\theta_{\vec{k}} +\!\!\left[ \beta    \delta^{\prime}( \tau ) - k^2 \right] \! w_{\vec{k} ||} \, .
\end{eqnarray}
Here $w_{\vec{k} ||}$ and $\vec{w}_{\vec{k}\perp}$ are respectively the longitudinal (i.e. parallel
to the wave-vector $\vec{k}$) and transversal (i.e. perpendicular to $\vec{k}$) components of the Fourier mode
$\vec{w}_{\vec{k}}(t)$, $\beta=2^{9/10} q_0\approx0.323$ and $\tilde{\mu}_1=1.811$ are the constants. To obtain Eqs. (\ref{Fmodes1})-(\ref{Fmodes4}) we keep only leading-order terms with respect to the dissipative parameter $\delta^{\, \prime} (t)$ in the expansion Eq. (\ref{etakapmu}) for the coefficients $\eta, \, \kappa, \, \mu, \, \zeta$ \cite{BrilliantovPoeschel:2003}. Note the important difference between the linearized equations (\ref{Fmodes1})-(\ref{Fmodes4}) and the corresponding  equations for the case $\varepsilon = {\rm const}$: In the latter case the coefficients in these equations are constant (e.g.
\cite{BreyRuizMonteroCubero:1999,NoijeErnst:2000}), while in the former case they depend on time via
\begin{equation}
\delta^{\prime}( \tau ) = 2^{1/10} \delta \left[ 1 + 2 q_0 \delta \, \tau /5 \right]^{-1} \, .
\end{equation}
The solution for $\vec{w}_{\vec{k}\perp} (\tau)$ reads
\begin{equation}
\vec{w}_{\vec{k}\perp} (\tau) = \vec{w}_{\vec{k}\perp} (0) \left[ 1+ 2 q_0 \delta \, \, \tau/5 \right]^5 e^{-k^2
\tau} \, ,
\label{eq:linstabperp}
\end{equation}
where $\vec{w}_{\vec{k}\perp} (0) $ is the initial amplitude of
the shear mode. There exists a critical wave-number, $k^*_{\perp}
\equiv \sqrt{2 q_0 \delta}$, which separates two regimes: Shear
modes with  $k \ge k^*_{\perp}$ always decay, while those with $k
< k^*_{\perp}$ initially grow and reach  a maximum,
\begin{equation}
w^{\rm max}_{\vec{k}\perp}= w_{\vec{k}\perp} (0) \left[\frac{2q_0 \delta} {k^2 \,e}\right]^5
\exp\left(\frac{5k^2}{2q_0 \delta}\right)
\end{equation}
at $\tau_{k \perp}^*=5/k^2-5/(2 q_0 \delta)$, then they decay and
die off completely. The formation of vortices is attributed to the
growth of the shear mode $\vec{w}_{\vec{k}\perp}$
\cite{BreyRuizMonteroCubero:1999,NoijeErnst:2000}, therefore, the
vortices of size $\sim k^{-1}$ decay after a transient time $\sim
\tau_{k \perp}^*$. Any system of size $\sim L$ has a minimal
wave-number $ \sim 2\pi/L$, hence {\em all} shear modes will decay
by the time $\tau_{\perp}^* \sim 5L^2/4 \pi^2 -5/(2q_0\delta)$.

To perform the stability analysis for the other three modes we write Eqs. (\ref{Fmodes2})-(\ref{Fmodes4}) in the form
\begin{equation}
\dot{{\bf \Psi}}_k=\hat{\bf M}_k(\tau){\bf \Psi}_k \, , \qquad {\bf \Psi}_k \equiv  \left(\rho_{\vec{k}},
\theta_{\vec{k}}, w_{\vec{k} ||} \right)^{T} \, .
\end{equation}
The matrix $\hat{\bf M}_k(\tau)$ has {\em time-dependent}
eigenvalues and eigenmodes, which for small  dissipation $\delta$
are analogous to the sound and the heat mode of a gas with a
constant $\varepsilon$
\cite{BreyRuizMonteroCubero:1999,NoijeErnst:2000}.  For large
wave-numbers $k$ all the modes decay, while for small $k$ the heat
mode may grow.  The critical $k$ may be found  from the condition
$\dot{{\bf \Psi}}_k=0$, or ${\rm det } |\hat{\bf M}_k(\tau) |=0$:

\begin{equation}
\label{kparal} k_{||}^*(\tau) \simeq  \frac{1}{4 \sqrt{5}} \left( 2^{1/10}\beta \delta \right)^{1/2} \left[ 1+
\frac 25  q_0 \delta \tau \right]^{-1/2}  \, .
\end{equation}
Modes with $k > k_{||}^*(0)$ always decay, while those with $k < k_{||}^*(0)$ may initially grow. Since $k_{||}^*(\tau)$ decreases with time even a mode with $k < k_{||}^*(0)$ which initially grows, starts  to decay after a transient time, when the condition $k > k_{||}^*(\tau)$ is fulfilled.

The value of $k_{||}^*(\tau)$ becomes smaller than the minimal
wave-number $2 \pi /L$ at time $\tau_{||}^* \sim L^2/16
-5/(2q_0\delta)$, i.e., for $\tau > \tau_{||}^*$ the amplitude of
any of the modes $\rho_{\vec{k}}$, $\theta_{\vec{k}}$, $w_{\vec{k}
||}$ decays. Note that $\tau_{\perp}^* > \tau_{||}^*$ in agreement
with the simulation results, Figs. \ref{MDHYDTrans},
\ref{HYDparden}. Since clustering and vortex formation is
attributed to the instability of the heat and shear modes
\cite{GoldhirschZanetti:1993,BritoErnst:1998,BreyDuftyKimSantos:1998}
we conclude that the eventual decay of all modes predicted by the
linear stability analysis, implies the transient structure
formation.

We have studied a force-free granular gas of viscoelastic
particles by means of MD, numerical solution of the HD equations
and linear stability analysis of the HD equations. All three
methods indicate that structure formation in a gas of viscoelastic
particles occurs only as a transient phenomenon, whose duration
increases with the system size $N$ for the same particle number
density. Correspondingly, for larger $N$ the structures appear to
be denser and more long-lived, whereas
 an extrapolation to the unbounded system cannot be easily
concluded from these arguments.
The finite duration of the cluster phase is in a sharp contrast to the case of a gas with a simplified collision model $\varepsilon = {\rm const}$,
where structures have been proven to arise and to continuously
develop. In our simulations, due to the limited system size and
periodic boundary conditions, we have unphysical self-interactions
of the clusters. We believe, however, that this effect does not
invalidate the main conclusion of our study of the transient
character of structure formation in force-free granular gases.

%1\\2\\3\\4\\5\\6\\7\\8\\9\\10\\1\\2\\3\\4\\5\\6\\7\\8\\9\\20\\1\\2


\begin{thebibliography}{99}
\bibitem{secpeak} For $ w_{k, \, ||}(t)$ we observed a second peak, Fig. \ref{HYDparden}. Presently, it is not clear, whether it is attributed only to nonlinear effects or to the periodic boundary conditions too.
\bibitem{remarkalpha}
The rate of clusters growth, $-\dot{n}$, is proportional to the
cluster-particle collision rate, which  is smaller than the
interparticle collision rate, $\sim n \tau_c^{-1} \sim
n^2T^{1/2}$. Hence the estimates $-\dot{n} < {\rm const} \, n^2
T^{1/2}$, $T\sim t^{-z}$ ($z>0$) yields $n>1/t^{(1-z/2)}$, i.e.
$\alpha <1$.
\end{thebibliography}
\end{document}